\documentclass[runningheads]{llncs}
\usepackage{graphicx}
\usepackage{amsmath}
\usepackage{comment}
\usepackage{epstopdf}
\usepackage{amsmath}
\usepackage{amsfonts}
\usepackage{bbm}
\usepackage{caption}
\usepackage{subcaption}
\newcommand{\rpm}{\raisebox{.2ex}{$\scriptstyle\pm$}}
\usepackage{hyperref}
\hypersetup{colorlinks,urlcolor=blue}
\begin{document}
\title{Unsupervised Domain Adaptation with Semantic Consistency across Heterogeneous Modalities for MRI Prostate Lesion Segmentation}
\titlerunning{Unsupervised Domain Adaptation with Semantic Consistency}
\author{Eleni Chiou \inst{1, 2}  %index{Chiou, Eleni} 
\and Francesco Giganti \inst{3, 4} %index{Giganti, Francesco}
\and Shonit Punwani  \inst{5} %index{Punwani, Shonit}
\and Iasonas Kokkinos \inst{2} %index{Kokkinos, Iasonas}
\and Eleftheria Panagiotaki \inst{1, 2}} %index{Panagiotaki, Eleftheria}

\authorrunning{Chiou et al.}
\institute{Centre for Medical Image Computing, UCL, London, UK
\and Department of Computer Science, UCL, London, UK 
\and Department of Radiology, UCLH NHS Foundation Trust, London, UK 
\and Division of Surgery \& Interventional Science, UCL, London, UK
\and Centre for Medical Imaging, Division of Medicine, UCL, London, UK
\email{eleni.chiou.17@ucl.ac.uk}}
\maketitle              % typeset the header of the contribution

\begin{abstract}
Any novel medical imaging modality that  differs from previous protocols e.g. in the number of imaging channels,
introduces a new domain that is heterogeneous from previous ones. This common medical imaging scenario is rarely considered in the domain adaptation literature, which handles shifts across domains of the same dimensionality. In our work we rely on stochastic generative modeling to translate across  two heterogeneous domains at pixel space and introduce two new loss functions that promote semantic consistency. Firstly, we introduce a semantic cycle-consistency loss in the source domain to ensure that the translation preserves the semantics. Secondly, we introduce a pseudo-labelling loss, where we translate target data to source, label them by a source-domain network, and use the generated pseudo-labels to supervise the target-domain network. Our results show that this allows us to extract systematically better representations for the target domain. In particular, we address the challenge of enhancing performance on VERDICT-MRI, an advanced diffusion-weighted imaging technique, by exploiting labeled mp-MRI data.  When compared to several unsupervised domain adaptation approaches, our approach yields substantial improvements, that consistently carry over to the semi-supervised and supervised learning settings.
 
\end{abstract}

\section{Introduction}
Domain adaptation  transfers knowledge from a label-rich `source' domain to a label-scarce or unlabeled `target' domain. This allows us to train robust models targeted towards novel medical imaging modalities or acquisition protocols with scarce supervision, if any. Recent unsupervised domain adaptation methods typically achieve this  either by minimizing the discrepancy of the feature and/or output space for the two domains or by learning a mapping between the two domains at the raw pixel space. Either way, these approaches usually consider moderate domain shifts where the dimensionality of the input feature-space between the source and the target domain is identical. 

\begin{figure}[!t]
\centering
\includegraphics[width=0.8\textwidth]{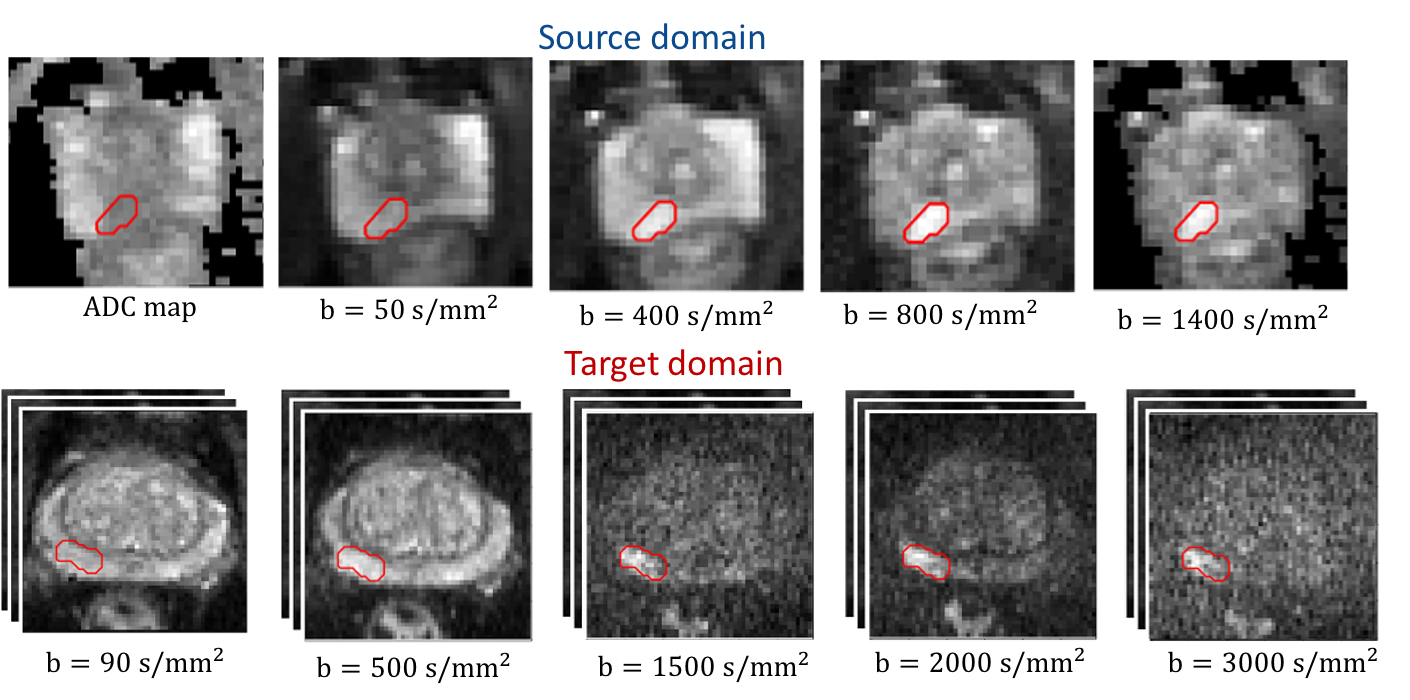}
\caption{Example of heterogeneous domains: standard DW-MRI (source domain) consists of 5 input channels (4 b-values and the ADC map) while VERDICT-MRI consists of 15 input channels (5 b-values in 3 orthogonal directions).}
\label{source_target_dom}
\end{figure}

In this work we address the challenge of adapting across two heterogeneous domains where both the distribution and the dimensionality of the input features are different (Fig.~\ref{source_target_dom}). Inspired by~\cite{harn_unc}, we rely on stochastic translation~\cite{Huang_ECCV_18} to align the two domains at pixel-level;~\cite{harn_unc} shows that stochastic translation yields clear improvements in heterogeneous domain adaptation tasks compared to deterministic, CycleGAN-based~\cite{CycleGAN} translation approaches. However, these improvements have been obtained with semi-supervised learning, where a few labeled target-domain images are available, whereas our goal is unsupervised domain adaptation. To this end we introduce a \textit{semantic cycle-consistency} loss on the cycle-reconstructed source images; if a source  image is translated to the target domain and then back to the source domain, we require that critical structures are preserved. We also introduce a \textit{pseudo-labeling} loss that allows us to use the unlabeled target data to  supervise the target-domain segmentation network. In particular we translate the target data to the source domain, predict their labels according to a pre-trained source-domain segmentation network and use the generated pseudo-labels to supervise the target-domain segmentation network. This allows us to use exclusively target-domain statistics and train highly discriminative models.

We demonstrate the effectiveness of our approach in prostate lesion segmentation and an advanced diffusion-weighted MRI (DW-MRI) method called VERDICT-MRI (Vascular, Extracellular and Restricted Diffusion for Cytometry in Tumours)~\cite{Verdict1,Verdict2}. Compared to the naive DW-MRI from multiparametric (mp)-MRI acquisitions, VERDICT-MRI has a richer acquisition protocol to probe the underlying microstructure and reveal changes in tissue features similar to histology. VERDICT-MRI has shown promising results in clinical settings, discriminating normal from malignant tissue~\cite{Verdict2,Chiou_ISMRM_19,Valindria_21} and identifying specific Gleason grades~\cite{Verdict3}.
However, the limited amount of available labeled training data does not allow us to utilize data-driven approaches which could directly exploit the information in the raw VERDICT-MRI~\cite{Chiou_MLMI_18,Chiou_ISMRM_20}. On the other hand, labeled, large scale clinical mp-MRI datasets exist~\cite{prostateX,promis}. In this work we exploit labeled mp-MRI data to train a segmentation network that performs well on VERDICT-MRI. 

\noindent\textbf{Related Work:}
Most domain adaptation methods  align the two domains either by extracting domain-invariant features or by aligning the two domains at the raw pixel space. Ren et al.~\cite{Ren_MICCAI_18} and Kamnitsas et al.~\cite{Kamnitsas_IPMI_17}, rely on adversarial training to align the feature distributions between the source and the target domain for medical image classification and segmentation respectively. Pixel-level approaches \cite{Jiang_MICCAI_18,Zhang_MICCAI_18,Cai_MedIA_19,Zhang_CVPR_18,harn_unc}, use GAN-based methods~\cite{CycleGAN,Huang_ECCV_18} to align the source and the target domains at pixel level. Chen et al.~\cite{synergistic} align simultaneously the two domains at pixel- and feature-level by utilizing adversarial training. Ouyang et al.~\cite{Quyang_MICCAI_19} combine a variational autoencoder (VAE)-based feature prior matching and pixel-level adversarial training to learn a domain-invariant latent space which is exploited during segmentation. Similarly,~\cite{Yang_MICCAI_19} perform pixel-level adversarial training  to extract content-only images and use them to train a segmentation model that operates well in both domains. Other studies exploit unlabeled target domain data during the discriminative training. Bateson et al.~\cite{source_relaxed} and Guodong et al. ~\cite{Guodong_2020} use entropy minimization on the prediction of target domain as an extra regularization while~\cite{dual_teacher} propose a teacher-student framework to train a model using labeled and unlabeled target data as well as labeled source data. 

\section{Method}
\subsection{Problem formulation}
We consider the problem of domain adaptation in prostate lesion segmentation. Let $\mathcal{X}_S \subset \mathbb{R}^{H\times W  \times C_S}$ be a set of $N_S$ source images and $\mathcal Y_S \subset \{0, 1 \}^{H, W}$ their segmentation masks. The sample $X_S \in \mathcal{X}_S$ is a $H\times W  \times C_S$ image and the entry $Y_S^{(h, w)}$ of the mask $Y_S$ provides the label of voxel $(h,w)$ as a one-hot vector. Let also $\mathcal{X}_T \subset \mathbb{R}^{H\times W  \times C_T}$ be a set of $N_T$ unlabeled target images. Sample $X_T \in \mathcal{X}_T$ is an $H\times W  \times C_T$ image. 

\begin{figure}[!t]
\centering
\includegraphics[width=0.9\textwidth]{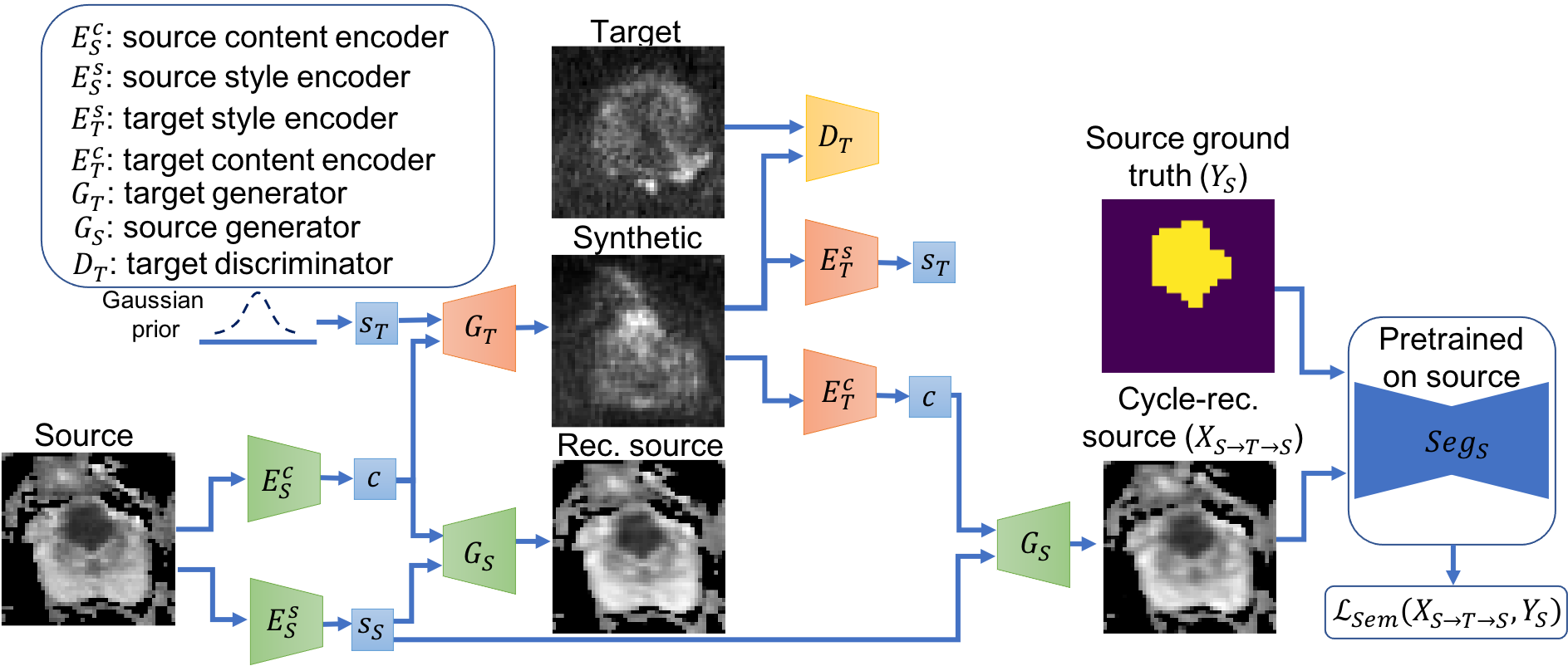}
\caption{We force a network for stochastic translation across domains to preserve semantics through a semantic segmentation-based loss. The image-to-image translation network translates source-domain images to the style of the target domain by combining a domain-invariant content code $c$ with a random code $s_T$. We introduce a semantic cycle-consistency loss, $\mathcal{L}_{Sem}$, on the cycle-reconstructed images that ensures that the prostate lesions are successfully preserved.}
\label{transl}
\end{figure}

\subsection*{Stochastic translation with semantic cycle-consistency regularization}
We rely on stochastic translation~\cite{harn_unc,Huang_ECCV_18} to learn the mapping  between the two domains and introduce a semantic cycle-consistency loss to enforce the cross-domain mapping to preserve critical structures.

The image-to-image translation network (Fig.\ref{transl}) consists of content encoders $E_S^c$, $E_T^c$, style encoders $E_S^s$, $E_T^s$, generators $G_S$, $G_T$ and domain discriminators $D_S$, $D_T$ for both domains. The content encoders $E_S^c$, $E_T^c$ extract a domain-invariant content code $c \in \mathcal{C}$ ($E_S^c: \mathcal{X}_S \rightarrow \mathcal{C}$, $E_T^c: \mathcal{X}_T \rightarrow \mathcal{C}$) while the style encoders $E_S^s$, $E_T^s$ extract domain-specific style codes $s_S \in \mathcal{S}_S$ ($E_S^s: \mathcal{X}_S \rightarrow \mathcal{S}_S$) and $s_T \in \mathcal{S}_T $ ($E_T^s: \mathcal{X}_T \rightarrow \mathcal{S}_T$). Image-to-image translation is performed by combining the content code ($c=E_S^c(X_S)$) extracted from a given input (${X_S \in \mathcal{X}_S}$) and a random style code $s_T$ sampled from the target-style space. We note  that the random style-code sampled from a Gaussian distribution represents structures that cannot be accounted by a deterministic mapping and results in one-to-many translation. 
We train the networks with a loss function consisting of domain adversarial, self-reconstruction, latent reconstruction and semantic cycle-consistency losses.

\noindent \textbf{Domain adversarial loss}. 

\noindent 

$\mathcal L_{GAN}^T =
 \mathbb{E}_{c_S \sim \mathcal C, s_T \sim \mathcal S_T}[\log(1-D_T(G_T(c_S, s_T)))] 
+ \mathbb{E}_{X_T \sim \mathcal X_T}[\log(D_T(X_T))]$.

\noindent \textbf{Self-reconstruction loss}.

$\mathcal L_{recon}^S = \mathbb{E}_{X_S \sim \mathcal X_S}[\|G_S(E_S^c(X_S), E_S^s(X_S))-X_S\|_1].$

\noindent \textbf{Latent reconstruction loss}. 

$\mathcal L_{recon}^{c_S} = \mathbb{E}_{X_S \sim \mathcal X_S, s_T \sim \mathcal S_T}[\|E_T^c(G_T(E_S^c(X_s), s_T))- E_S^c(X_s)\|_1].
$ 

$\mathcal L_{recon}^{S_T} = 
\mathbb{E}_{X_S \sim \mathcal X_S, s_T \sim \mathcal S_T} [\|E_T^s(G_T(E_S^c(X_s), s_T))- s_T)\|_1].
$ 

\noindent \textbf{Cycle-consistency loss}. 

$
\mathcal L_{cyc}^{S} = 
\mathbb{E}_{X_S \sim \mathcal X_S, s_T \sim \mathcal S_T} [\|G_S(E_T^c(G_T(E_S^c(X_S), s_T)), E_S^s(X_S))-X_S\|_1].
$

\noindent $\mathcal L_{GAN}^S$, $\mathcal L_{recon}^T$, $\mathcal L_{recon}^{c_T}$, $\mathcal L_{recon}^{s_S}$, $\mathcal L_{cyc}^{T}$ are defined in a similar way. 

\noindent \textbf{Semantic cycle-consistency loss}. Recent studies~\cite{Jiang_MICCAI_18,Cai_MedIA_19,Zhang_CVPR_18,harn_unc} enforce semantic consistency between the real source and the synthetic target images by exploiting a target-domain segmentation network trained on a few available labeled target-domain images. However, in the unsupervised scenario, where there is no supervision available for the target domain, such approach is not feasible. To this end we introduce a semantic cycle-consistency loss or lesion segmentation loss on the cycle-reconstructed source images $X_{S \rightarrow T \rightarrow S}$; if a source  image is translated to the target domain and then back to the source domain, we require that critical structures, corresponding to lesions, are preserved. The naive cycle-consistency loss, introduced in~\cite{CycleGAN}, penalizes inconsistencies in the entire image  and may fail to preserve small structures corresponding to lesions. In contrast our semantic cycle-consistency loss penalizes inconsistencies in the label space enforcing the translation network to preserve the lesions. The semantic cycle-consistency loss is a soft generalization of the dice score given by
\begin{equation}
\mathcal L_{Sem} = - \frac {2 \sum_{h, w} P^{(h,w,1)} Y_S^{(h,w,1)}} {\sum_{h, w} (P^{(h,w,1)} + Y_S^{(h,w,1)})},
\end{equation}
where $P^{(h,w,1)}$ is the predictive probability of class 1 for voxel $(h,w)$ provided by the pre-trained source network $Seg_S$.
The full objective is given by
\begin{equation}
\begin{aligned}
& \min_{\substack{E_S^c, E_S^s, E_T^c, E_T^s, G_S, G_T}} \max_{ D_S, D_T} 
 \lambda_{GAN} (\mathcal L_{GAN}^S + \mathcal L_{GAN}^T) + \lambda_{x} (\mathcal L_{recon}^S + \mathcal L_{recon}^T) \\
&\quad \quad \quad +\lambda_{c} (\mathcal L_{recon}^{c_S} + \mathcal L_{recon}^{c_T})
+\lambda_{s} (\mathcal L_{recon}^{s_S} + \mathcal L_{recon}^{s_T}) + \lambda_{cyc} (\mathcal L_{cyc}^{S} + \mathcal L_{cyc}^{T}) \\
&\quad \quad \quad +\lambda_{sem} \mathcal L_{sem},
\end{aligned}
\end{equation}
where $\lambda_{GAN}$, $\lambda_{x}$, $\lambda_{c}$, $\lambda_{s}$, $\lambda_{cyc}$, $\lambda_{sem}$ control the importance of each term.

\begin{figure}[!t]
\centering
\includegraphics[width=0.9\textwidth]{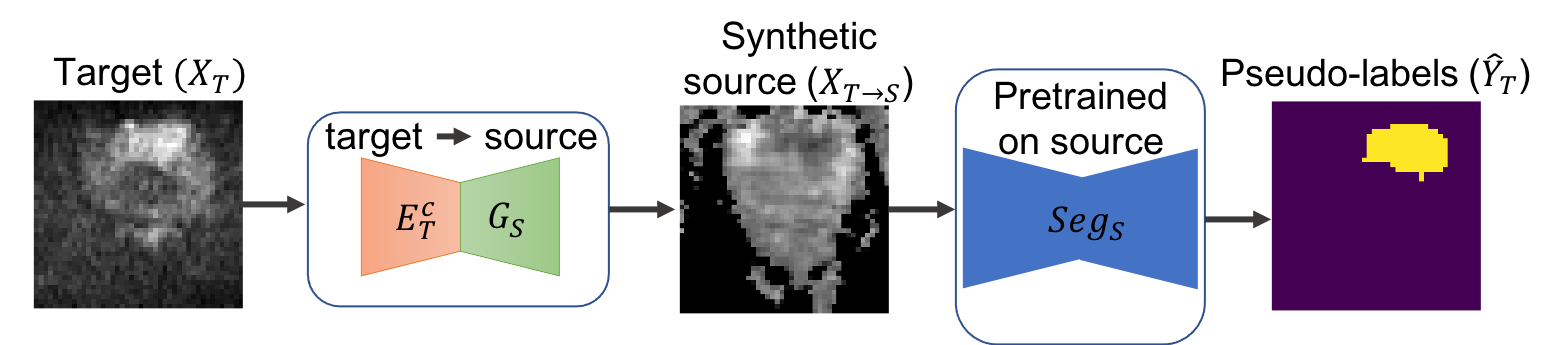}
\caption{Pseudo-labeling through translation to  the source domain: We translate the target data to the source domain and predict their pseudo-labels according to a pre-trained source-domain segmentation network $Seg_S$.}
\label{pseudo}
\end{figure}
\subsection*{Pseudo-labeling through translation to the source}
We generate pseudo-labels for the target images by translating them to the source domain and predicting their labels according to the pre-trained source-domain segmentation network $Seg_S$, trained on the labeled source data (Fig.~\ref{pseudo}).

Given a synthetic source image $X_{T \rightarrow S}$ and the segmentation network $Seg_S$ we obtain a soft-segmentation map 
$P_{X_{T \rightarrow S}} = Seg_S(X_{T \rightarrow S})$, where each vector $P_{X_{T \rightarrow S}}^{(h,w)}$ corresponds to a probability distribution over classes. 
Assuming that high-scoring pixel-wise predictions on synthetic source samples are correct, we obtain a segmentation mask $\hat{Y}_T$ by selecting high-scoring pixels with a fixed threshold. Each entry $\hat{Y}_T^{(h,w)}$ can be either a discrete one-hot vector for high-scoring pixels or a zero-vector for low-scoring pixels. The pseudo-labeling configuration is defined as follows
\begin{equation}
\hat{Y}_T^{(h,w,c)} = \begin{cases} 1,  &\mbox{if } c = \underset{c}{\arg\max} P_{X_{T \rightarrow S}}^{(h,w)} ~ \rm{and} ~ P_{X_{T \rightarrow S}}^{(h,w,c)} > \rm{threshold} \\
0, & \mbox{otherwise.} \end{cases}
\end{equation}

\begin{figure}[!t]
\centering
\includegraphics[width=0.9\textwidth]{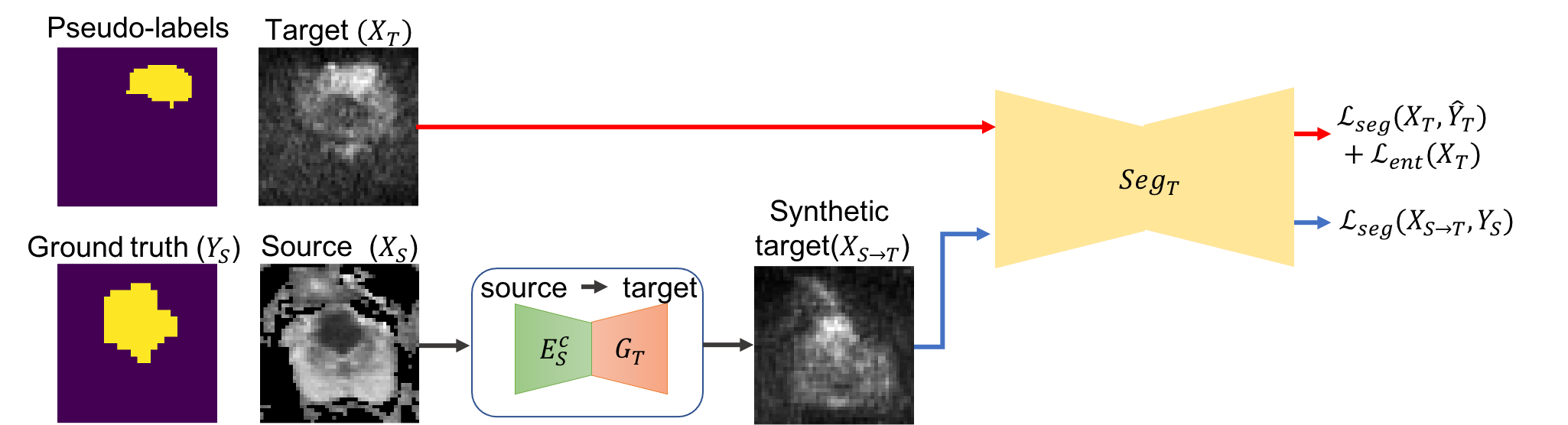}
\caption{We use data that have exclusively target-domain statistics to train the target segmentation network ($Seg_T$). We translate the source data to the target domain and supervise $Seg_T$ using the ground-truth segmentation masks. We also use target pseudo-labels to supervise $Seg_T$.}
\label{seg}
\end{figure}

\subsection*{Segmentation Network}
The target-domain segmentation network, $Seg_T$, is an  encoder-decoder network~\cite{Chen_ECCV_18,Unet}. We supervise $Seg_T$ using both the synthetic target images and the corresponding source labels and the real target images and their pseudo-labels (Fig.~\ref{seg}).
Given an image $X$ and its segmentation mask $Y$, the segmentation loss is defined as 
\begin{equation}
\mathcal L_{Seg}(X, Y) = - \frac {2 \sum_{h,w} P^{(h,w,1)} Y^{(h,w,1)}} {\sum_{h,w} (P^{(h,w,1)} + Y^{(h,w,1)})},
\end{equation}
where $P^{(h,w,1)}$ is the predictive probability of class 1 for voxel $(h,w)$. 

As in recent studies \cite{Advent,source_relaxed}, to further regularize the network on the target-domain data for which we have not obtained pseudo-labels, we apply entropy-based regularization. The loss $\mathcal L_{ent}$ is defined as follows
\begin{equation}
\mathcal L_{Ent}(X_T) = \sum_{h,w} \frac{-1}{\log{C}} \sum_{c=1}^C P_{X_{T}}^{(h,w,c)} \log {P_{X_{T}}^{(h,w,c)}}
\end{equation}
\noindent The full objective is given by
$
\min_{\substack{Seg_T}}\mathcal \mathcal L_{Seg} + \mathcal L_{Ent}.
$

\subsection{Implementation details}
We implement our framework using Pytorch~\cite{pytorch}. Our code is publicly available at \url{https://github.com/elchiou/SemanticConsistUDA}.

\noindent \textbf{Image-to-image  translation network}: The content encoders consist of convolutional 
layers and residual blocks followed by instance normalization \cite{IN}. The style encoders consist of convolutional layers followed by fully connected layers. The decoders include residual blocks followed by upsampling and convolutional layers. The residual blocks are followed by adaptive instance normalization (AdaIN)~\cite{AdaIN} layers to adjust the style of the output image. The discriminators consist of convolutional layers. For training we use Adam optimizer, a batch size of 32, a learning rate of 0.0001  and set losses weights to $\lambda_{GAN}=1$, $\lambda_{x}=10$, $\lambda_{c}=1$, $\lambda_{s}=1$, $\lambda_{sem}=10$. We train the translation network for $50000$ iterations. 
\noindent \textbf{Segmentation network}: The encoder of the segmentation network is a standard ResNet~\cite{resnet} consisting of convolutional layers while the decoder consists of upsampling and convolutional layers. For training we use stochastic gradient decent and a batch size of 32. We split the training set into $80 \%$ training and $20 \%$ validation to select the learning rate, the number of iterations and the threshold to perform pseudo-labeling. 

\section{Datasets}
\textbf{VERDICT-MRI}: We use VERDICT-MRI data  from $60$ men. The DW-MRI was acquired with pulsed-gradient spin-echo sequence (PGSE) and an optimized imaging protocol for VERDICT prostate characterization with 5 b-values ($90$, $500$, $1500$, $2000$, $3000 \ \rm{s/mm^2}$) in 3 orthogonal directions~\cite{acq_prot}. The DW-MRI sequence was acquired with a voxel size of $1.25 \times 1.25 \times 5 \ \rm{mm^3}$, $5 \ \rm{mm}$ slice thickness, $14$ slices, and field of view of $220 \times 220 \ \rm{mm^2}$ and the images were reconstructed to a $176 \times 176$ matrix size. A radiologist contoured the lesions on VERDICT-MRI using mp-MRI for guidance.

\noindent\textbf{DW-MRI from mp-MRI acquisition}: We use DW-MRI data from $80$ patients from the ProstateX challenge dataset~\cite{prostateX}. Three b-values were acquired ($50, 400, 800  \  \rm{s/mm^2}$), and the ADC map and a b-value image at $\rm{b} = 1400\  \rm{s/mm^2}$ were calculated by the scanner. The DW-MRI data were acquired with a single-shot echo planar imaging sequence with a voxel size of $2 \times 2 \times 3.6\ \rm{mm^3}$, $3.6\  \rm{mm}$ slice thickness. Since the ProstateX dataset provides only the position of the lesion, a radiologist manually annotated the lesions on the ADC map using as reference the provided position.

\begin{table}[!t]
\center
\caption{Average recall, precision, dice similarity coefficient (DSC), and average precision (AP) across 5 folds. The results are given in mean (\rpm std) format.} \label{table}
\resizebox{1\textwidth}{!}{
\begin{tabular}{|l|l|l|l|l|}
\hline
Model                                                           & Recall         & Precision              & DSC          & AP                     \\ \hline
VERDICT-MRI (Oracle)                                            & 66.2 (8.1)     & 70.5 (9.9)             & 68.9 (9.2)   & 72.1 (10.4)            \\ \hline  
VERDICT-MRI + Synth (MUNIT + $\mathcal L_{Sem}$)                & 71.1 (8.9)     & 72.5 (10.4)            & 72.1 (8.7)   & 76.7 (9.6)             \\ \hline \hline
mp-MRI + EntMin (ADVENT~\cite{Advent})                          & 50.8 (12.3)    & 48.0 (11.4)            & 49.8 (13.0)  & 51.4 (13.9)            \\ \hline
Synth (MUNIT)                                                   & 51.5 (13.3)    & 60.6 (11.9)            & 53.6 (12.7)  & 60.2 (13.0)            \\ \hline
Synth (MUNIT + $\mathcal L_{{Sem}}$)                            & 55.1 (13.9)    & 62.4 (12.8)            & 55.3 (10.9)  & 62.0 (13.4)            \\ \hline
Synth (MUNIT + $\mathcal L_{Sem}$) + EntMin (Ours)              & 54.7 (11.5)    & \textbf{69.2} (10.3)   & 57.1 (10.8)  & 63.4 (12.8)            \\ \hline
Synth (MUNIT + $\mathcal L_{Sem}$) + PsLab (Ours)               & 59.8 (10.1)    & 64.8 (11.1)            & 61.5 (10.3)   & 64.9 (10.1)            \\ \hline
Synth (MUNIT + $\mathcal L_{Sem}$) + EntMin + PsLab (Ours)      & \textbf{61.4} (9.9) & 66.9 (10.7)       & \textbf{62.1} (9.8)   & \textbf{65.6} (10.9)   \\ \hline
\end{tabular}}
\end{table}

\section{Results}

\begin{figure}[!t]
\centering
\begin{minipage}{.45\textwidth}
\centering
\includegraphics[width=1\textwidth]{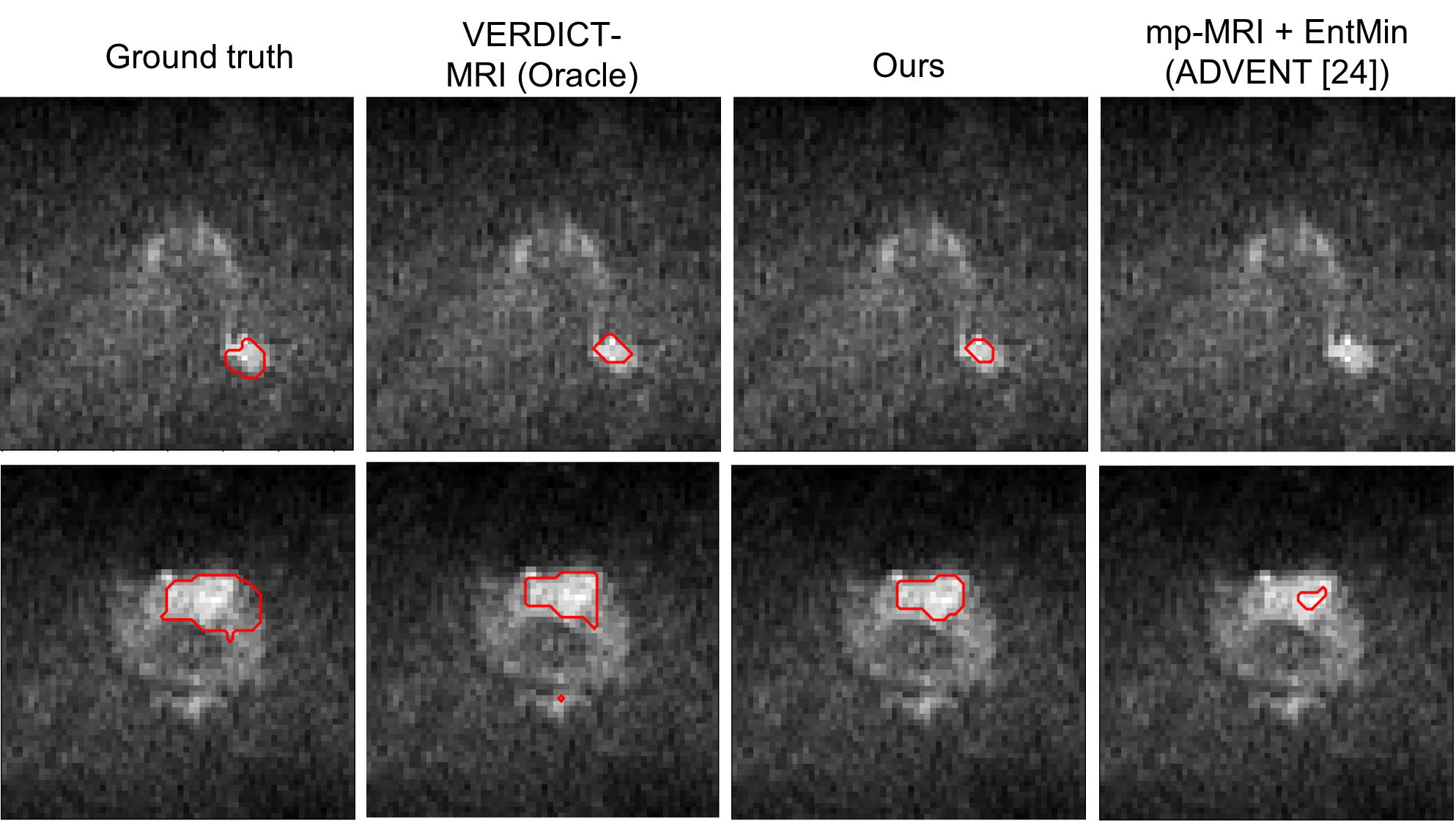}
\caption{Lesion segmentation results for two patients - the proposed approach  performs well on the target domain despite the fact that it does not utilize any manual target annotations during training.}
\label{qual_res}
\end{minipage}
\hspace{0.5cm} 
\begin{minipage}{.45\textwidth}
\centering
\includegraphics[width=1\textwidth]{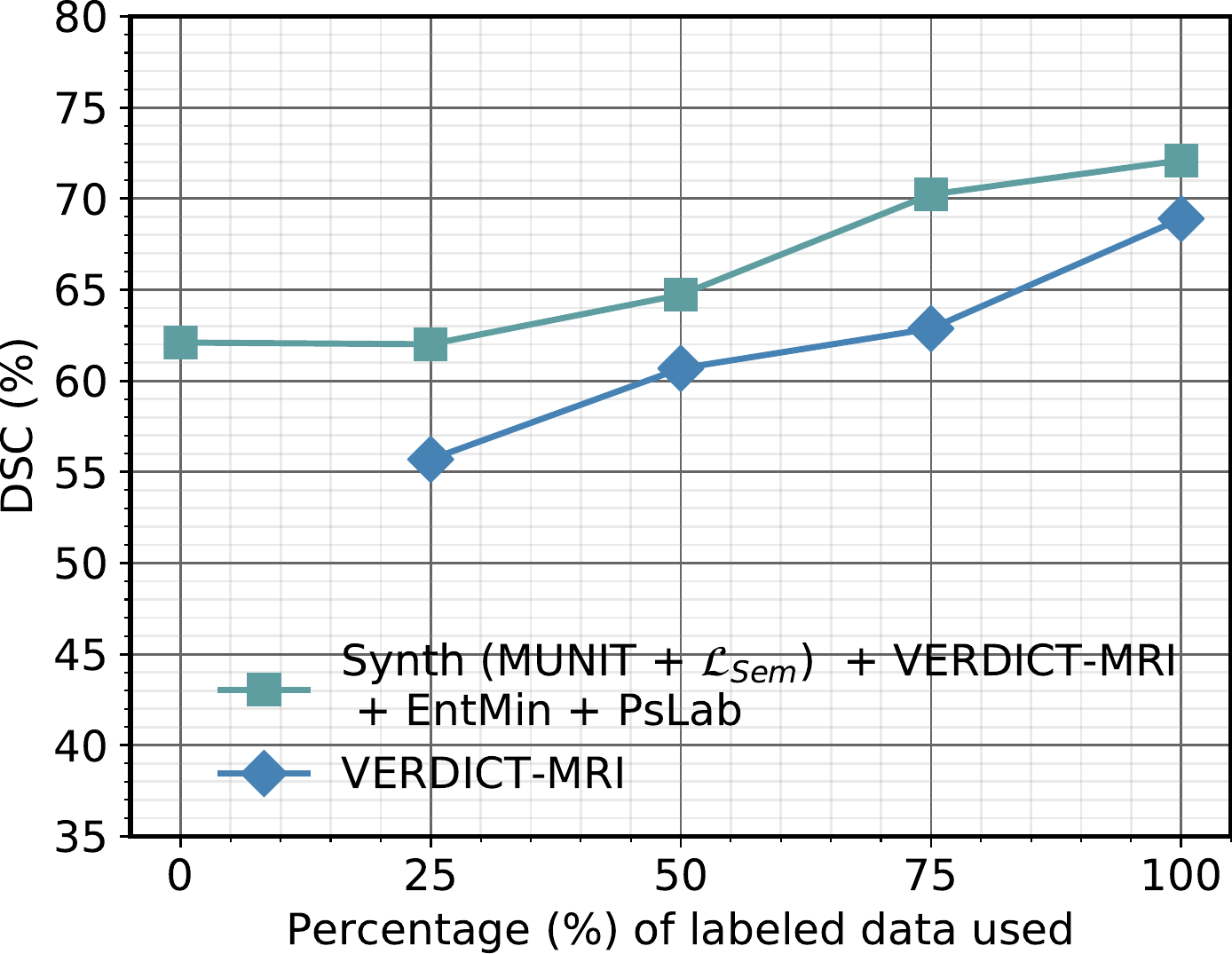}
\caption{Performance as we vary the percentage of labeled target data used for training. We observe that our method improves with more supervision and the improvements introduced by our method over the baseline of target-only training carry over all the way to the fully-supervised regime.}
\label{perc_real}
\end{minipage}
\end{figure} 
We evaluate the performance based on the average recall, precision, dice similarity coefficient (DSC), and average precision (AP) across 5-folds.

We compare our approach to several baselines. i)VERDICT-MRI: train using VERDICT-MRI only. ii) VERDICT-MRI + Synth (MUNIT + $\mathcal L_{Sem}$)~: train using real VERDICT-MRI and the synthetic VERDICT-MRI obtained from MUNIT with semantic cycle-consistency loss. iii) mp-MRI + EntMin (ADVENT~\cite{Advent}): train the model by minimizing the segmentation loss, $\mathcal{L}_{Seg}(X_S, Y_S)$, on the raw mp-MRI and the entropy loss, $\mathcal{L}_{Ent}(X_T)$, on VERDICT-MRI, an approach proposed in~\cite{Advent,source_relaxed,Guodong_2020}.  iv) Synth (MUNIT): use the naive MUNIT to map from source to target and train only on the synthetic data. v) Synth (MUNIT + $\mathcal L_{{Sem}}$): use MUNIT with semantic cycle-consistency loss to translate from source to target. vi) Synth (MUNIT + $\mathcal L_{{Sem}}$) + EntMin: use (v) and entropy-based regularization on VERDICT-MRI data. 
vii) Synth (MUNIT + $\mathcal L_{{Sem}}$)  + PsLab: use (v) and pseudo-labels to train the segmentation network on real VERDICT-MRI. viii) Synth (MUNIT + $\mathcal L_{{Sem}}$) + EntMin + PsLab: use (vi) and pseudo-labels to train the segmentation network on real VERDICT-MRI. 

We report the results in Table \ref{table}. We observe that the performance is poor when the segmentation network is trained  on the mp-MRI and VERDICT-MRI data (mp-MRI + EntMin (ADVENT~\cite{Advent})). However, we observe that when we train the network with synthetic VERDICT-MRI and real VERDICT-MRI (Synth (MUNIT + $\mathcal L_{{Sem}}$) + EntMin) the performance improves. This justifies our assumption that pixel-level alignment is beneficial in cases where there is a large distribution shift. The performance further improves when we use pseudo-labels obtained from confident predictions (Synth (MUNIT + $\mathcal L_{{Sem}}$) + EntMin + PsLab). We also observe that compared to the naive MUNIT without the semantic cycle-consistency loss (Synth (MUNIT)) our approach (Synth (MUNIT + $\mathcal L_{{Sem}}$)) performs better since it successfully preserves the lesions. When combining real and synthetic data (VERDICT-MRI + Synth (MUNIT + $\mathcal L_{Sem}$)) to train the network in a fully-supervised manner we get better results compared to the oracle, where we use only the real VERDICT-MRI. In Figure~\ref{qual_res} we present lesion segmentation results produced by the different models for two patients. The results indicate that the proposed approach performs well despite the fact that it does not use any manual annotations during training.

So far we have considered only the unsupervised case. However, our approach can also be used in a semi-supervised setting. To evaluate the performance of our method when labeled target data is available, we perform additional experiments varying the percentage of labeled data; we use the pseudo-labels (PsLab) and entropy minimization (EntMin) for the unlabeled data. Figure~\ref{perc_real} shows that the performance of our method improves as the percentage of real data increases and always outperforms the baseline that is trained only on the target domain.

\section{Conclusion}
In this work we propose a domain adaptation approach for lesion segmentation. Our approach relies on appearance alignment along with pseudo-labeling to train a target domain classifier using exclusively target domain statistics. We demonstrate the effectiveness of our approach for lesion segmentation on VERDICT-MRI which is an advanced imaging technique for cancer characterization. However, the proposed work is a general method that can be extended to other applications where there is a large distribution gap between the source and the target domain. 
\bibliographystyle{splncs04}
\bibliography{bibliography}

\end{document}